# Efficient FIR filtering with Bit Layer Multiply Accumulator

Vincenzo Liguori        Ocean Logic Pty Ltd        enzo@ocean-logic.com

**Abstract**
Bit Layer Multiplier Accumulator (BLMAC) is an efficient method to perform dot products without multiplications that exploits the bit level sparsity of the weights. A total of 1,980,000 low, high, band pass and band stop type I FIR filters were generated by systematically sweeping through the cut off frequencies and by varying the number of taps from 55 to 255. After their coefficients were quantized to 16 bits, applying the filter using a BLMAC required, on average, from ~123.3 to ~513.6 additions, depending on the number of taps. A BLMAC dot product machine, specialised for 127 taps FIR filters, was designed for AMD FPGAs. The design footprint is ~110 LUTs, including coefficient and sample storage and is able to apply the filter in ~232 clock cycles on average. This implies a filtering rate of 1.4-3.4 Msamples/s, depending on the FPGA family.

# 1. Introduction

Bit Layer Multiplier Accumulator (BLMAC) is an efficient method to perform dot products without multiplications by exploiting the bit level sparsity of the weights. Such sparsity lends itself naturally to their compression that can be, optionally, integral part of the BLMAC framework.

This paper focuses on BLMAC performance on a specific application of dot product, FIR filters. This is done by determining the average number of additions required by a BLMAC on a sample of approximately 2 million type I FIR filters.

Application of BLMAC to Convolutional Neural Networks (CNNs) can be found in [1].

The paper starts with an introduction to BLMAC and how it compares to existing techniques for dot product calculations. This has mostly already been published in [1]. However, here there's some additional discussion about the BLMAC precision.

Next the generation of the ~2 million FIR filters is outlined as well the quantisation of their coefficients to 16 bits. This is followed by an analysis of the performance of the BLMAC when applying the filters. Finally, a BLMAC dot product machine design, specialised for 127 taps FIR filters, is introduced and discussed.

Albeit the emphasis of this paper is efficient hardware implementation, the techniques described here could be applied in software to low end processors that do not have efficient multiply accumulate (MAC) instructions.

# 2. Bit Layer Multiply Accumulator

Lets' consider the definition of dot products between a weight vector $\vec{w}$ and an input vector $\vec{x}$ of dimensionality N:

$$\vec{w} \cdot \vec{x} = \sum_{j=0}^{N-1} w_j x_j \quad (1)$$

This can be calculated with N multiplications and N-1 additions. Typical architectures to calculate this are Multiply Accumulators (MACs), shown in Fig.1a) and Fig.1d) for the serial version.

In order to simplify the discussion, we will start with assuming the weights to be positive integers. Negative weights will be dealt with later.

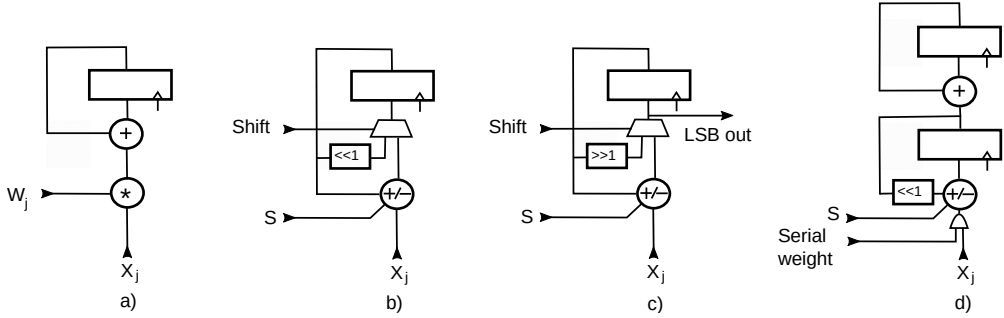

Figure 1 Dot product hardware architectures.

Each component $w_j$ of $\vec{w}$ is a positive integer and it can represented as a binary $w_j$ number $w_j = \sum_{i=0}^{N_b-1} d_{ij} 2^i$ with $d_{ij} \in \{0,1\}$ :

$$\vec{w} \cdot \vec{x} = \sum_{j=0}^{N-1} \vec{w}_j x_j = \sum_{j=0}^{N-1} \left( \sum_{i=0}^{N_b-1} d_{ij} 2^i \right) x_j = \sum_{i=0}^{N_b-1} \left( \sum_{j=0}^{N-1} d_{ij} x_j \right) 2^i =$$

$$(\ldots((\sum_{j=0}^{N-1} d_{N_b-1\,j} x_j) 2 + \sum_{j=0}^{N-1} d_{N_b-2\,j} x_j) 2 + \ldots) 2 + \sum_{j=0}^{N-1} d_{0j} x_j \quad (2)$$

Where N is the dimensionality of the vectors and $N_b$ the number of bits of the weights. Eq. 2 shows that $\vec{w} \cdot \vec{x}$ can be calculated by starting with the contribution $\sum_{j=0}^{N-1} d_{N_b-1\,j} x_j$ of all the most significant bits (MSB) of the weights. We then multiply by 2 and add the contribution of the second MSB. We continue to do that until we reach the least significant bit (LSB). If we look at the matrix $d_{ij}$ of binary digits we are performing multiply and accumulate operations from the MSB row to the LSB one, one bit layer at the time. Hence the name bit layer MAC (BLMAC). We will indicate non-zero elements in a bit layer ($d_{ij} \neq 0$) as "pulses".

Let's consider a simple example where $\vec{w} = (1, 27, 7, 0, 2)$ and $\vec{x} = (x_0, x_1, x_2, x_3, x_4)$ , Nb=N=5. Then $\vec{w} \cdot \vec{x} = x_0 + 27 x_1 + 7 x_2 + 2 x_4$ . We can calculate the same using equation 2.

These vectors are shown on Tab.1 with a column containing the binary representation of a weight.

| | | | | | |
|---|---|---|---|---|---|
| *4* | 0 | 1 | 0 | 0 | 0 |
| *3* | 0 | 1 | 0 | 0 | 0 |
| *2* | 0 | 0 | 1 | 0 | 0 |
| *1* | 0 | 1 | 1 | 0 | 1 |
| *0* | 1 | 1 | 1 | 0 | 0 |
| | 1 | 27 | 7 | 0 | 2 |
| | $x_0$ | $x_1$ | $x_2$ | $x_3$ | $x_4$ |

Table 1: Bit Layer MAC example.

We start from row or layer 4, the MSB layer. With only one pulse, we get $x_1$. We now multiply by 2 and continue to add the contribution of layer 3: $2x_1 + x_1 = 3x_1$. Again we multiply by 2 and add the contribution of layer 2: $6x_1 + x_2$. We continue with layer 1: $2(6x_1 + x_2) + x_1 + x_2 + x_4 = 13x_1 + 3x_2 + x_4$. Finally, with layer 0: $2(13x_1 + 3x_2 + x_4) + x_0 + x_1 + x_2 = x_0 + 27x_1 + 7x_2 + 2x_4$. Which is the expected result.

Fig.1c shows one of the possible architecture for a BLMAC. The S input controls the add/subtract behaviour of the accumulator. Since this is only relevant to negative weights, we will disregard this input for now and only assume addition capabilities. Multiplying a value by 2 in binary is a simple shift to the left, indicated by <<1.

Starting from the MSB layer i=$N_b$-1, we present at the input all the values $x_j$ for which $d_{ij}$≠0 which will be added to the accumulator every clock cycle. Once all the pulses in a layer are exhausted the `shift` input will cause the accumulator to shift one position to the left. We then continue to process all the pulses in the next layer and so on, until all the layers are processed. We note that :

- A BLMAC is essentially an add/sub accumulator plus a 2:1 multiplexer to select the shift operation.
- In many cases (like in CNNs), the dimensionality N of a dot product is much larger than the number of bits $N_b$ of a weight (a few thousands vs. no more than a dozen). Therefore the $N_b$ shifts required by the BLMAC can be considered negligible. Even for FIR filter the overhead is small. Besides, the architectures in Fig.1b could be modified to perform the shift in the same clock cycle.
- If we skip all the zero $d_{ij}$ values, then the number of additions needed to calculate $\vec{w}\cdot\vec{x}$ is equal to the total number of pulses in all the weights. This is analogous to a sparse weight situation with "weights" are now either 0 or 1 (0 and +/-1 with negative numbers).
- The smaller the number of pulses, the faster is to calculate $\vec{w}\cdot\vec{x}$ . This architecture exploits sparsity at bit level, not just at weight level.
- A BLMAC is naturally a variable precision MAC: it can deal with different size weights efficiently, without the need of sizing the design according to the worst case (largest weight). The same architecture will work with binary, integer and even floating point weights. The latter might appear surprising until one considers that a fp32 number can be seen as a very large but bit sparse integer (up to a scaling factor) with, at most, 24 non zero bits.
- The design in Fig.1b is based on Eq. 6. It starts from the MSB bit layer and uses left shifts. The same result is obtained by starting from the LSB and using right shifts. With each right shift one bit of the final result is output. This version can have some advantages over the other, such as smaller adder and accumulator as the bits that are right shifted out are no longer part of the computation. This will be referred to as "right shift BLMAC", shown in Fig.1c.

We now come to negative weights. The simplest way is to encode a weight as two's complement number. The same design in Fig.1b can be used. Incoming $x_j$ values will be subtracted from the accumulator when processing the sign layer (with S input = 1). For all the other bit layers, $x_j$ values will be added (as before).

This is because, in two's complement representation, the sign bit is a negative power of two. For example, if we have a 3 bit numbers and we want add a sign to it, the value -1 will be $1111_2$ = $-2^3+2^2+2^1+2^0$.

There is a problem with this representation, however. It requires a lot of pulses (think -1=$1111_2$ vs. 1=$0001_2$. This will have a negative impact on the processing speed as more pulses require more clock cycles.

This can be solved if we pass to a ternary representation with $d_{ij} \in \{-1,0,1\}$ but still represent weights as

$$w_j = \sum_{i=0}^{N_b-1} d_{ij} 2^i$$

. Now positive and negative numbers use exactly the same number of pulses. For example, if we look at number 5, $101_2$ in binary, then -5 will be (-1,0,-1)=$-2^2-2^0$.

Moreover, we can now reduce the number of pulses in some cases. For example $11111_2$ can also be represented by (1,0,0,0,0,-1): 2 pulses instead of 5. Another case is $11011_2$ = (1,0,0,-1,0,-1): 3 pulses instead of 4. There are many other similar cases and this representation can help reduce the number of pulses and hence increase the speed. We can now repeat the same example of Tab. 1 with the new encoding.

| 5 | 0 | 1 | 0 | 0 | 0 |
|---|---|---|---|---|---|
| 4 | 0 | 0 | 0 | 0 | 0 |
| 3 | 0 | 0 | 1 | 0 | 0 |
| 2 | 0 | -1 | 0 | 0 | 0 |
| 1 | 0 | 0 | 0 | 0 | 1 |
| 0 | 1 | -1 | -1 | 0 | 0 |
|   | 1 | 27 | 7 | 0 | 2 |
|   | $x_0$ | $x_1$ | $x_2$ | $x_3$ | $x_4$ |

Table 2: Same BLMAC example with new encoding.

Starting from row 5 and get $x_1$. Next row there are no pulses, so we just shift : $2x_1$. At row 3 we shift again and add $x_2$, obtaining $4x_1+x_2$. At row 2 we shift and subtract x1 : $8x_1+2x_2-x_1 = 7x_1+2x_2$. At row 1 we shift and add x4 : $14x_1+4x_2+x_4$. Finally comes row 0, we shift and add/subtract the contributions according to the weights: $2(14x_1+4x_2+x_4)+x_0-x_1-x_2 = x_0+28x_1-x_1+8x_2-x_2+2x_4 = x_0+27x_1+7x_2+2x_4$.

The design of Fig.1c can still be used. This time, when a pulse is positive, $x_j$ values will be added, subtracted otherwise. The polarity of the pulse will control the S input.

It is also worthwhile to show how the same result is also obtained by starting from the LSB layer and shift right at the end of each bit layer. Every time this architecture shifts the accumulator right, a bit of the result is output. Here, in order to show the full result, all the inputs will be scaled by $32=2^5$ (5 is the number of bit layers).

So, starting from the LSB layer 0 : $32x_0-32x_1-32x_2$. Next layer 1, right shift : $16x_0-16x_1-16x_2$ and add $32x_4$. Result is $16x_0-16x_1-16x_2+32x_4$. Layer 2, shift and subtract $32x_1$ : $8x_0-8x_1-8x_2+16x_4-32x_1 = 8x_0-40x_1-8x_2+16x_4$. Layer 3, shift and add $32x_2$ : $4x_0-20x_1-4x_2+8x_4+32x_2 = 4x_0-20x_1+28x_2+8x_4$. Layer 4 is just a shift : $2x_0-10x_1+14x_2+4x_4$. Finally layer 5, shift and add $32x_1$ : $x_0-5x_1+7x_2+2x_4+32x_1 = x_0+27x_1+7x_2+2x_4$.

## 2.1. BLMAC Precision

Calculating a dot product can result in a large number of bits [2]. For example, in the case of FIR filters with 16 bit coefficients, filtering 8 bit samples, 24 bits are potentially needed for the product. We then add them multiple times. For a 255 tap filter this could require up to 32 bits.

Ideally, even if one doesn't need all the bits in the final result, for maximum precision, the result should only be truncated/rounded at the end of the computation. There's always the option to truncate/round the multiplication result before accumulating it, but that will reduce the precision of the result.

However, a right shift BLMAC has a useful property: at the end of the calculation of the $i^{th}$ bit layer, the LSB of the accumulator contains the $i^{th}$ bit of the final result. This bit is fully determined and it won't be affected by any further bit layer calculations and this means that it can be potentially be thrown out when shifting. This allows to manage the precision of the final result without the consequences of the truncated multiplication.

Even in the unlikely case that one needs all the bits of the resulting dot product, the BLMAC adder is smaller as it does not have to operate on the bits already shifted out. In fact, whatever the number of bits of the final result, we know that, since $n_b$ bits are shifted out during the computation, the adder in the BLMAC is $n_b - 1$ bits smaller than a MAC's. And being smaller, it's also faster.

## 2.2. BLMAC is NOT a Serial MAC

Although the operation that BLMAC and serial MAC (i.e. a MAC that performs the multiplication serially and then adds the result to an accumulator (Fig.1d)) perform is the same, it is important to stress that organising the computation along bit layers confers inherent advantages to the former:

- **Size**. In a serial MAC, two registers are required: one for the multiplication by shifts and adds and one to add the result to the accumulator. Two adders are also needed, although a single one could potentially be shared. This is at least twice the size of a BLMAC. Also, as mentioned above, the size of the adder in a right shift BLMAC is $n_b - 1$ bits smaller.

- **Speed**. Another problem is due to the shifts. For example, if we serially multiply $x_j$ by $10001_2$, we start with $x_j$, we shift it 4 times to the left and then add $x_j$ again. If these shifts are performed serially, this will take 5 cycles versus 2 in a BLMAC. This is because a BLMAC only perform a shift at the end of each bit layer instead of every single weight. Adding a barrel shifter could allow multiple shifts in the same clock cycle. This again adds complexity. Moreover, when adding a barrel shifter one would need to decide what is the maximum number of simultaneous shifts. More shift means larger logic, mostly underutilised with infrequent large weights. Few shifts means additional cycles and lower performance in some cases. This not a problem for BLMAC that handle variable precision naturally without costly barrel shifters.

- **Precision**. See previous section.

In conclusion, a BLMAC is substantially different, smaller, faster and potentially more precise than a serial MAC.

## 2.3. BLMAC Performance

The table below shows, for integers up to 24 bits, the average and the maximum number of pulses. Column 7, for example, contains these statistics for all the 7 bit integers (i.e. all the integers between 0 and +127 included).

| $N_b$ | 1 | 2 | 3 | 4 | 5 | 6 | 7 | 8 | 9 | 10 | 11 | 12 | 13 | 14 | 15 | 16 | 17 | 18 | 19 | 20 | 21 | 22 | 23 | 24 |
|---|---|---|---|---|---|---|---|---|---|---|---|---|---|---|---|---|---|---|---|---|---|---|---|---|
| **Avg** | 0.5 | 1.0 | 1.37 | 1.75 | 2.09 | 2.44 | 2.77 | 3.11 | 3.44 | 3.77 | 4.11 | 4.44 | 4.78 | 5.11 | 5.44 | 5.77 | 6.11 | 6.44 | 6.78 | 7.11 | 7.44 | .7.78 | 8.11 | 8.44 |
| **Max** | 1 | 2 | 2 | 3 | 3 | 4 | 4 | 5 | 5 | 6 | 6 | 7 | 7 | 8 | 8 | 9 | 9 | 10 | 10 | 11 | 11 | 12 | 12 | 13 |

Table 3: Average and maximum number of pulses for numbers of given bit size.

So column 7 tells us that any integer in such range can be encoded with a maximum of 4 pulses. This is also the maximum number of cycles that a BLMAC will take to multiply by a weight in the same range and, if the weights are uniformly distributed, we can expect to take an average of ~2.77 cycles each.

Again, for an N-dimensional dot product with 7 bits weights that are uniformly distributed, one can expect the BLMAC to take (on average) ~2.77*N additions or cycles instead of the normally required N multiplication and N-1 additions.

A negative number takes exactly the same number of cycles, as the only difference is the inverted pulse sign. Thus column 7 also includes the case of 8 bits signed numbers.

This values provided here are only valid for uniformly distributed weights. This is often not the case for practical applications. For example, CNN usually have a skewed distribution of the weights with many weights and fewer larger ones [3]. In this case BLMAC is likely to have better performance than the one shown in Tab. 3, as can be seen here [1].

As we shall see, this is also true for FIR filters.

## 2.4. BLMAC Dot Product Machines

Computing dot products along sparse bit layers lends itself naturally to weight compression. Each bit layer can be represented as run-lenght pairs (S,ZRUN) where S is the +/-1 value of a non zero dij coefficient and ZRUN is the number of zero values preceding it.

When the non zero coefficients in a bit layer are exhausted, this can be indicated by an End Of Run code (EOR). An empty bit layer is simply indicated by an EOR code. For example, layer 0 of Tab.2 can be represented by the codes (1,0),(-1,0),(-1,0), EOR.

These code can be (optionally) compressed further using entropy encoding. This brings us to the architecture of a BLMAC dot product machine shown in Fig.2.

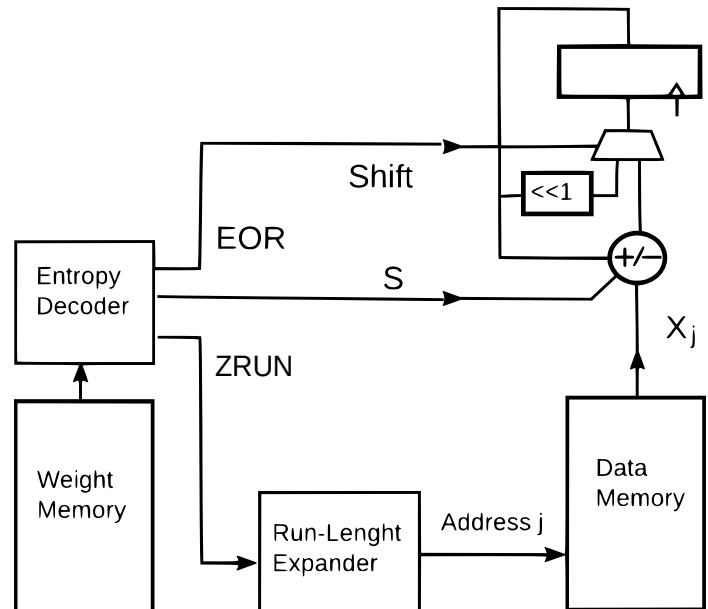

Figure 2 A possible architecture for a BLMAC dot product machine.

At the start of a new dot product the BLMAC accumulator is cleared. Then, for each bit layer, a (S,ZRUN) pair is read and decoded from the compressed memory. The run-lenght code is then expanded, generating and index j that is used to retrieve $x_j$ from the data memory.

This is then added or subtracted from the BLMAC accumulator according to the decoded S. When a EOR code is decoded, the BLMAC performs a shift. When all the codes are exhausted, the accumulator contains the dot product result.

A parallel version of this architecture was used in a CNN accelerator [1].

# 3. BLMAC Performance on FIR Filters

This section discusses a performance test of the BLMAC on a sample of some 2 million type I FIR filters. This is broken up in how they are generated, quantized and evaluated.

## 3.1. Generating the FIR Filters

The type I FIR filters are generated using the `firwin` function from the Python `scipy` library by systematically sweeping through the cut off frequencies and varying the number of taps between 55 and 255.

The cut off frequency in `firwin`, by default, is twice the normalised Nyquist frequency and must be 0.0 < f < 1.0. So, for low pass filters, if we divide the frequency space in N parts, we will generate N-1 filters. For example, with N = 10, the cut off frequencies will be 0.1, 0.2, …., 0.9. The same is done for high pass filters.

For band pass filters is a bit more complicated as we need two cut off frequencies $f_1$ and $f_2$ such that 0.0 < $f_1$ < $f_2$ < 1.0. Again, if we divide the frequency space in N parts, we will generate (N-1)*(N-2)/2 filters. For example [0.1, 0.2], [0.1,0.3],….,[0.1,  0.9], then [0.2, 0.3], [0.2, 0.4],….,[0.2, 0.9] etc. The same is done for band stop filters.

So, the total number of filters generated will be 2*(N-1) + 2*(N-1)*(N-2)/2 = N*(N-1). N was arbitrarily chosen to be 100, resulting in 9,900 filters.

Doing the same for all the taps from 55 to 255 (odd values only for type I FIR filters) gives another factor of 100, bringing the total to 990,000 filters.

The `firwin` function uses by default the Hamming window. So, another 990,000 FIR filters were generated with the Kaiser window. Therefore the total number of FIR filters generated is close to two millions.

## 3.2. Quantazing the Filters' Coefficients

The `firwin` function generates an array of N double precision floating point numbers for an N tap filter. These were quantized to 16 bits signed numbers.

In order to do that, the simplest thing would be to scale each coefficient by $2^{15}$ and round the result.

However, for many generated filters, the largest coefficients can be quite a bit smaller than 1.0. This means that, when quantized this way, the full available range of a signed 16 bits number will not be utilised.

Consequently, each coefficient was scaled by largest power of two such that the largest coefficient would still fit in a signed 16 bits word, see [2]. The result was then rounded with convergent rounding.

This was considered a fairer test for BLMAC as the quantized coefficients truly required 16 bits.

## 3.3. Performance and Analysis

When applying an N taps type I FIR filter, a common optimisation consists in exploiting the symmetry of the coefficients by adding the corresponding samples first.

$$\vec{w}\cdot\vec{x}=\sum_{j=0}^{N-1} w_j x_j = w_{N/2} x_{N/2} + \sum_{j=0}^{N/2-1} w_j (x_j + x_{N-j-1}) \quad (3)$$

We basically go from an N dimensional dot product to an N/2 + 1 dimensional dot product plus N/2 additions (where "/" is the integer division). This reduces the number of operations from the original N multiplications and N – 1 additions to N/2 + 1 multiplications and N – 1 additions.

We can use the same algorithm for BLMAC, pre-adding the samples and then perform an N/2 + 1 dimensional dot product. The performance here is measured by the total number of additions required. Therefore, in this case, it will be N/2 additions plus however many additions are taken by the BLMAC to compute the N/2 + 1 dimensional dot product.

In order to know how many additions the latter would take, we need to add the number of trits/pulses in the ternary representation of each coefficient. This can be done with a pre-calculated array `ntrits[]` containing, for all the possible $2^{15}$ coefficient values (not $2^{16}$ as the sign doesn't change the number of additions for a BLMAC), the number of trits/pulses. For example, `ntrits[118]` = 3 because 118 = (1,0,0,0,-1,0,-1,0), 3 non zero trits.

Given a vector `w[]` of first N/2+1 FIR coefficients we can count all the trits/additions required for applying the filter with a BLMAC :

`tot = N/2; // Count the preliminary additions`

`for(i=0;i<N/2+1;i++)`

   `tot += ntrits[abs(w[i])]; // Count the BLMAC contribution`

The results are shown in the graphs below for both the filters generated with Hamming (Fig.3) and the Kaiser (Fig.4) windows. For each number of taps from 55 to 255 the total number of additions required (`tot`) is averaged over the 9,900 filters with that particular number of taps and shown in blue. The standard deviation is shown as error bars. It comb-like appearance is due to the fact that type I FIR filters only have an odd number of taps. The plots also show in red the largest number of the additions as well as the smallest (in green) encountered while processing the 9,900 filters.

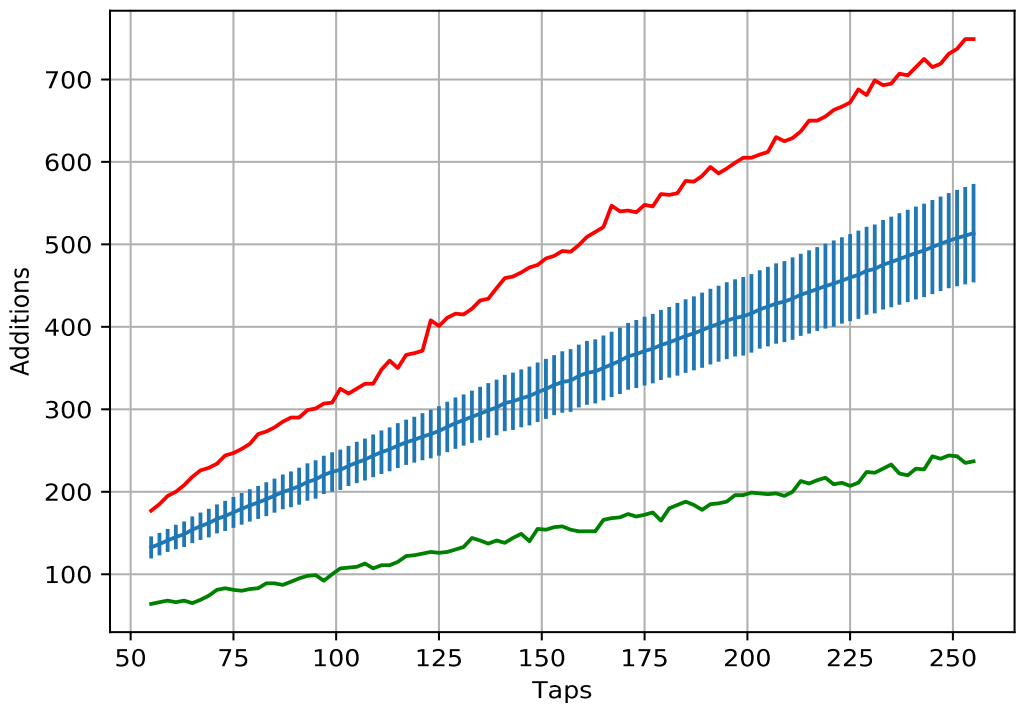

Figure 3 Number of additions for applying the FIR filters generated with the Hamming window.

For each number of taps N, we now have have the average number of additions required by BLMAC ($B_N$). $B_N$ varies from ~132.5 for N = 55 to ~513.6 for N = 255 for FIR filters generated with the Hamming window. For the Kaiser window ones, $B_N$ varies from ~123.3 for N = 55 to ~474.7 for N = 255.

This is to be compared to the N/2 + 1 multiplication plus N – 1 additions required by the classical algorithm.

Comparing additions to additions and multiplications is a bit of an apples and pears comparison but we could say that a 16 bits signed multiplication is really equivalent to 15 additions. Then the total number of additions for the classical algorithm is 15*(N/2+1) + N -1. So, for, say a 255 taps filter, we have 15*(255/2+1) + 255 -1 = 15*128 + 254 = 2,174 equivalent number of additions. However $B_{255}$ = ~513.6 for the Hamming windows is ~4.23 times better. Things are even better for the Kaiser window with $B_{255}$ = ~474.7.

We can also evaluate the performance of the BLMAC on the N/2 + 1 dimensional dot product by subtracting the N/2 preliminary additions from $B_N$. Then dividing $B_N$ - N/2 by the N/2 + 1 number of points will give us the average number of additions per coefficient when performing dot products with FIR coefficients.

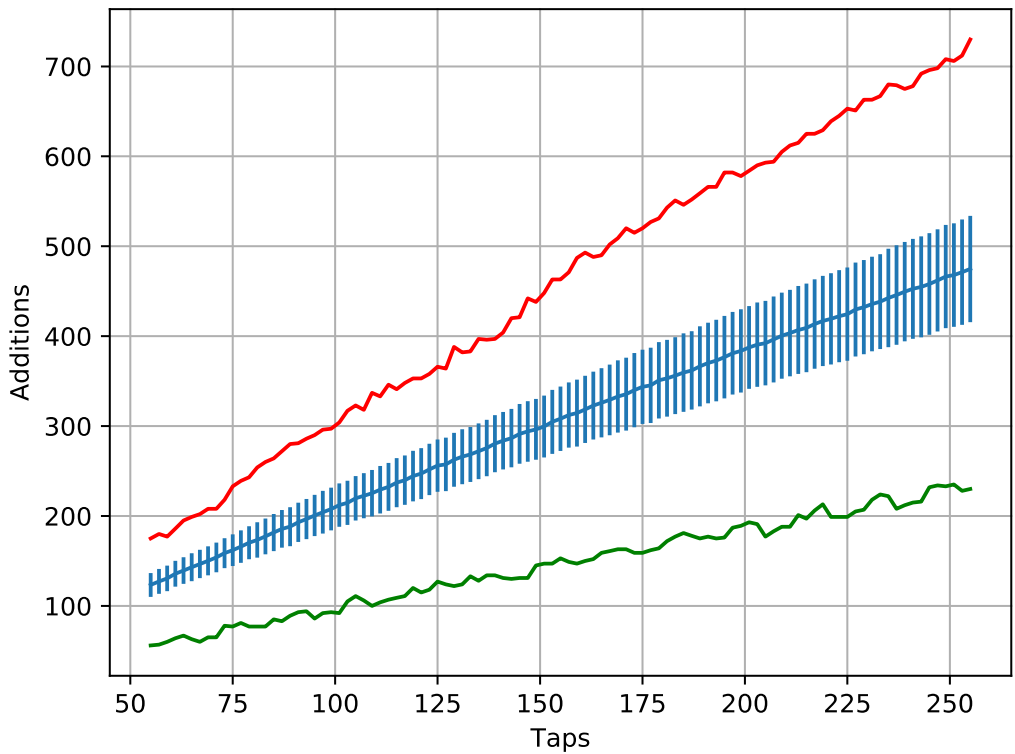

Figure 4 Number of additions for applying the FIR filters generated with the Kaiser window.

This number can be now directly compared to column 15 on Tab.3. So, $(B_N - N/2)/(N/2 + 1)$ varies from ~3.8 for N = 55 to ~3.0 for N = 255. This represents a ~40-80% improvement over the ~5.44 additions per coefficient that we could expect if they were uniformly distributed. Again, for the Kaiser window case, things are somewhat better with the ratio $(B_N - N/2)/(N/2 + 1)$ varying from ~3.4 for N = 55 to ~2.7 for N = 255.

This should come as no surprise to anyone that has ever looked at a plot of the coefficients of an FIR filter : on average, there are few large values and many smaller ones.

We can also simply consider the ratio $B_N/N$ : the number of additions required per tap. This ratio varies from ~2.41 to ~1.86 in these tests, decreasing with increasing N: only a few additions per tap.

One final note: here we have used the number of additions as a performance measure. However, as mentioned in section 2.1. and, as we shall see in section 4., the adder required is significantly smaller than one in an ordinary MAC. That should be kept in mind.

# 4. A 127 Taps FIR BLMAC Dot Product Machine

A BLMAC dot product machine, specialised for 127 taps FIR filters, was designed in AMD FPGA technology.

A simplified diagram for this prototype design is shown in Fig.5.

This design is a variation of the one in Fig.2 and it works in a similar way, taking into account the optimisation outlined in Eq. 3. A right shift BLMAC was used.

Before it's started, the weight memory is written with run length codes as described in section 2.4. A total of 127 8 bit samples is also stored in the sample memory. The BLMAC dot product machine is then started. Run length codes are read sequentially from the weight memory. Each run-lenght is expanded and used to read symmetrical samples from the sample memory. These are added before being added/subtracted according to the sign in the code.

Each time an EOR is encountered, a bit layer is completed and the accumulator is shifted. Every bit shifted out is collected in a shift register. After all the bit layers have been processed, the accumulator and the shift register contain the full result of applying the filter to the 127 samples.

We can now shift a new sample in the sample memory and the process is repeated again.

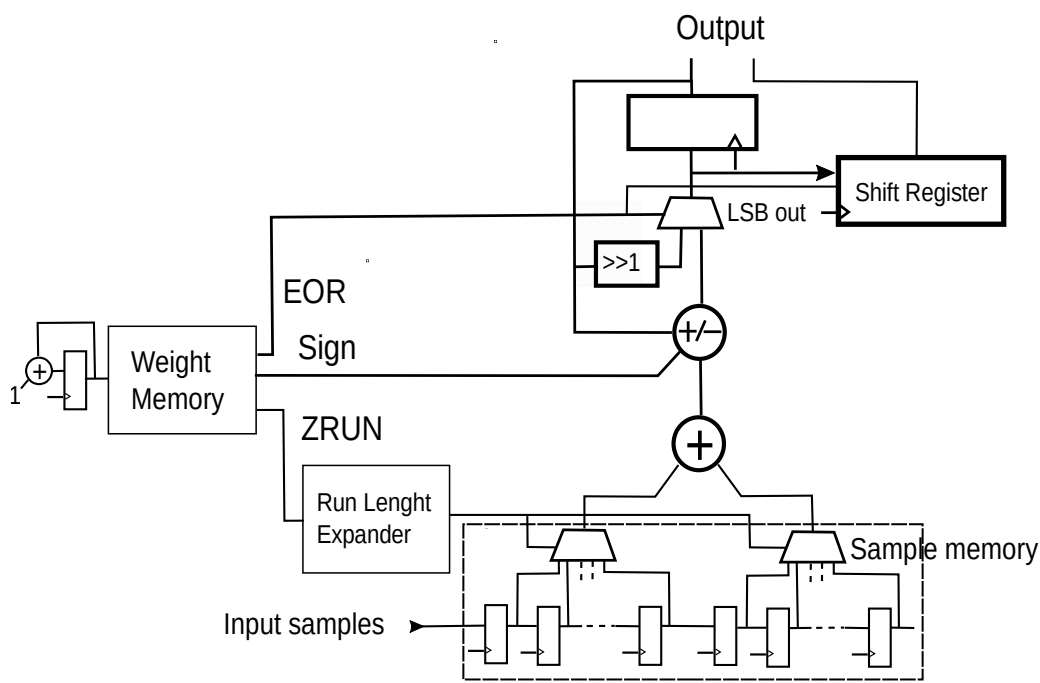

Figure 5: BLMAC dot product machine specialised for 127 taps FIR filter.

The design was tested on 9,900 127 taps filters generated with the Hamming window. A total of 126 + 256 random input input samples were generated for each filter and used by a C model to generated 256 expected output samples with the classical algorithm. No bits were truncated or rounded in the expected output. For each FIR filter, the run-lenght codes were also generated.

A testbench was constructed that, for each FIR filter, reads the run-lenght codes and programs the weight memory. It then writes the first 127 samples in the sample memory and starts the dot product machine. The result was checked for each of the 256 expected outputs for each filter.

When applying the various 127 tap filters, the design takes an average of ~231.6 clock cycles. The weight memory was only able to store up to 256 run-lenght codes and so, ~18% of the 9,900 filters could not be tested in this particular prototype design. However, the design is quite flexible and it can easily be modified for a different/variable number of taps as well as different size weight and sample memory.

Note that, by changing the pre-adder into an adder/subtractor, we could also apply type II, III and IV FIR filters. Another improvement would be to perform the last addition/subtraction at the end of the a bit layer at the same time as the shift. This would reduce the number of clock cycles by 16.

Tab.4 shows the design implemented in various families of AMD FPGAs with devices with the highest speed rate, optimized for area and speed.

|  | Area | | | Speed | | |
| --- | --- | --- | --- | --- | --- | --- |
| Family | LUTs | Max Freq (MHz) | Rate (Msample/s) | LUTs | Max Freq (MHz) | Rate (Msample/s) |
| Artix Ultrascale + | 117 | >800.0 | >~3.45 | 134 | > 800.0 | >~3.45 |
| Kintex Ultrascale + | 116 | >800.0 | >~3.45 | 134 | > 800.0 | >~3.45 |
| Artix 7 | 100 | 316.8 | ~1.37 | 134 | 416.1 | ~1.8 |
| Kintex 7 | 101 | 407.3 | ~1.76 | 134 | 628.5 | ~2.71 |

*Table 4 Implementation figures for AMD FPGAs.*

The rate indicates the average filtering rate, obtained by dividing the max clock frequency by the average number of clock cycles taken by the dot product machine.

Performance for the Ultrascale+ family were beyond 800 MHz, whether that's realistic or not.

The weight memory is implemented as a 256x8 distributed memory. The sample memory was also implemented as distributed memory. The BLMAC itself plus the 16 bits shift register is only 25 LUTs when synthesised in isolation. In fact, in most of not all practical cases the full result is not needed and the shift register could be eliminated, shaving off a few more LUTs from an already small design.

# 5. Conclusion

Although is not possible to completely explore the FIR filter space, this paper has shown, through its nearly 2 million samples, the effectiveness of the BLMAC for FIR filters application where, on average, only a few additions per tap are required. A simple, compact and fully programmable architecture for a specialised BLMAC dot product machine was also presented that is capable of applying FIR filters at mega samples per second rates without DSP primitives and with a small number of LUTs.

# References

[1] V. Liguori, "A MAC-less Neural Inference Processor Supporting Compressed, Variable Precision Weights", arXiv: 2012.06018 , 10 Dec 2020
[2] Randy Yates, "Practical Considerations in Fixed-Point FIR Filter Implementations", 11 Sep 2010, http://www.digitalsignallabs.com/fir.pdf
[3] S. Han, J. Pool, J. Tran, W.J.Dally "Learning both Weights and Connections for Efficient Neural Networks", arXiv : 1506.02626v3, 30 Oct 2015.